\documentclass[journal,comsoc]{IEEEtran}
\usepackage[T1]{fontenc}

\usepackage{amsmath}
\usepackage{paralist}
\usepackage{tabularx}
\usepackage{graphicx}
\usepackage{sidecap}
\usepackage{epstopdf}
\usepackage{amsthm}
\usepackage{textcomp}
 \usepackage{cite}
\usepackage{ragged2e}

\ifCLASSINFOpdf
\else
\fi
\usepackage{amsmath}
\usepackage[cmintegrals]{newtxmath}
\usepackage{algorithm}
\usepackage[noend]{algpseudocode}

\begin{document}
%
\title{Joint User Association and Resource Allocation Optimization for Ultra Reliable Low Latency HetNets} 

%
%
%

\author{Mohammad~Yousefvand,~\IEEEmembership{Member,~IEEE,}
        Narayan~B.~Mandayam,~\IEEEmembership{Fellow,~IEEE,} 

\thanks{ 
This work was supported in part by NSF under Grant No. 1421961 and Grant No. ACI-1541069 .
}

\thanks{ M. Yousefvand is with Wireless Information Network Laboratory (WINLAB, Rutgers University, NJ 08902.
E-mail: my342@winlab.rutgers.edu.}

\thanks{ NB. Mandayam is with Wireless Information Network Laboratory (WINLAB, Rutgers University, NJ 08902.
E-mail: narayan@winlab.rutgers.edu.}
}
%
%

\markboth{IEEE Transactions on Green Communications and Networking \LaTeX\ Class Files,~Vol.~, No.~, Sep~2018}%
{Shell \MakeLowercase{\textit{et al.}}: Bare Demo of IEEEtran.cls for IEEE Communications Society Journals}
%



\maketitle

\begin{abstract}
Ensuring ultra-reliable and low latency communications (URLLC) is necessary for enabling delay critical applications in 5G HetNets. We propose a joint user to BS association and resource optimization method that is attractive for URLLC in HetNets with Cellular Base Stations (CBSs) and Small Cell Base Stations (SBSs), while also reducing energy and bandwidth consumption. In our scheme, CBSs share portions of the available spectrum with SBSs, and they in exchange, provide data service to the users in their coverage area. We first show that the CBSs optimal resource allocation (ORA) problem is NP-hard and computationally intractable for large number of users. Then, to reduce its time complexity, we propose a relaxed heuristic method (RHM) which breaks down the original ORA problem into a heuristic user association (HUA) algorithm and a convex resource allocation (CRA) optimization problem. Simulation results show that the proposed heuristic method decreases the time complexity of finding the optimal solution for CBS's significantly, thereby benefiting URLLC. It also helps the CBSs to save energy by offloading users to SBSs. In our simulations, the spectrum access delay for cellular users is reduced by 93\% and the energy consumption is reduced by 33\%, while maintaining the full service rate.
\end{abstract}


%
\IEEEpeerreviewmaketitle

\section{Introduction}
\IEEEPARstart{R}{educing} transmit power and bandwidth consumption of base stations are crucial to enhance the energy and spectral efficiency of cellular networks, specially in high load situations where the user demands exceed available network resources \cite{Hajisami:2017:ENB}. In HetNets, it is well known that cellular base stations (CBSs) can save both spectrum and energy by offloading users to overlaid Small cell Base Stations (SBSs). 
To motivate SBSs to serve cellular users, the CBS grants some portion of its licensed spectrum to SBSs who serve the offloaded users, and the amount of granted bandwidth must be larger than that required to serve the offloaded users. After offloading, each CBS has to optimize its bandwidth and power allocation to those users who are retained and need to be served directly. Besides reducing the overall transmit energy and bandwidth consumption from the CBS point of view, specifically for URLLC we need to reduce the spectrum access delay for the users to enable delay sensitive applications. To do so, we need to find new resource allocation mechanisms for BSs that have a low time complexity besides being energy and bandwidth efficient.\\

In this work we consider a HetNet model with one CBS and multiple SBSs, and show that the CBS's Optimal Resource Allocation (ORA) problem, which includes the joint user association and bandwidth/power allocation problems, is a non-convex NP-hard problem. It is actually a combinatorial problem which is composed of three sub problems. The first sub-problem is the users-to-BSs association problem which assigns users to the CBS or SBSs and can be seen as a classification problem. The second and third sub-problems are power and bandwidth allocation problems, respectively. Then, to make the original ORA problem solvable in a reasonable time for a HetNet with a large number of users, we propose a new Relaxed Heuristic Method (RHM) which includes a heuristic user association algorithm, and a convex optimization problem for power/bandwidth allocation. After solving the user association problem using the proposed heuristic, and removing user association optimization variables from the original ORA problem, it reduces to a convex optimization problem that minimizes the CBS's power consumption, while serving users with the minimum bandwidth possible. To further reduce the time complexity of RHM and reduce the spectrum access delay, we use the Alternating Direction Method of Multipliers (ADMM) to solve the resource allocation optimization problem formulated in RHM. Simulation results show that our proposed heuristic method can significantly reduce the time complexity of finding the optimal solution for user association and resource allocation optimization problems in ORA. \\
The rest of this paper is organized as follows. Section \ref{sec2} provides the literature review, and section \ref{sec3} introduces the system and network model. The ORA problem is formulated in section \ref{sec4}, and the heuristic method is described in section \ref{sec5}. Simulation results are presented in section \ref{sec6}, and we conclude in section \ref{sec7}.

\section{Literature Review}
\label{sec2}
In wireless HetNets, user association, power allocation and spectrum allocation are tightly coupled as has been shown in the vast amount of literature \cite{Liu:2015:JUA, Hossain:2014:RAC, Han:2017:BAU, Borst:2013:ORA, Lee:2018:MPS, Sokun:2017:NOQ, Munir:2017:ROM, Ye:2016:UAI, Oo:2015:TOU, Amiri:2018:MLA, Niknam:2016:MHN}.
Most of these works are focusing on optimizing one of the BS resources while doing the user association. For example, in \cite{Liu:2015:JUA} a joint user association and green energy allocation scheme is proposed for HetNets with hybrid energy sources to minimize the on-grid energy consumption. A general framework for joint BS association and power control for wireless HetNets is presented in \cite{Hossain:2014:RAC}, in which a price-based non-cooperative game theory approach is taken to design a dynamic spectrum sharing algorithm for two-tier macro-femto networks. A distributed backhaul-aware user association and resource allocation method is proposed in \cite{Han:2017:BAU} for energy-constrained HetNets, where the authors used the Lagrange dual decomposition method to solve the user association problem, and showed that for the optimal resource allocation a BS either assigns equal normalized resources or provides an equal long-term service rate to its served users. The authors in \cite{Borst:2013:ORA} investigated the optimal user association and resource allocation problem for a network scenario with one macro cell and several non-interfering overlaid pico cells, and proposed an algorithm for achieving a max-min fair throughput allocation across all users. A self-organizing approach to joint user association and resource blanking between network-tiers in highly dense HetNets is developed in \cite{Lee:2018:MPS} where resource blanking is used to partition time slots into two orthogonal groups and dedicate them to macrocells and small cells exclusively using a novel message-passing algorithm. In \cite{Sokun:2017:NOQ} a combinatorial optimization problem is formulated for joint optimization of user to base station association, and time-frequency resource block and power allocation. Then, the authors provided an approximate solution for it using a polynomial complexity two phase approach based on semidefinite relaxation with randomization. In \cite{Munir:2017:ROM} the authors investigated the impacts of multi-slope path loss models, where different link distances are characterized by different path loss exponents, and then proposed a framework for joint user association, power and subcarrier allocation on the downlink of a HetNet. A novel framework for the joint optimization of user association and interference management in massive MIMO HetNets is presented in \cite{Ye:2016:UAI}, in which resource allocation problem is cast as a network utility maximization (NUM) problem, and then a dual subgradient based algorithm is proposed which converges toward the NUM solution. In \cite{Oo:2015:TOU} a distributed game model is presented for joint user association and resource allocation in HetNets. In \cite{Amiri:2018:MLA} a Q-learning based approach is proposed for adaptive resource optimization in multi-agent networks to satisfy users QoS demands and provide fairness among them. In \cite{Niknam:2016:MHN} the authors proposed a resource allocation scheme for the HetNets with one macro BS and several mmWave small cell BSs to maximize the sum rate of the network while ensuring the minimum rate requirement for each user.\\ 

However, most of these works propose combinatorial optimization problems with NP-hard complexities which cannot be solved in a reasonable time for URLLC applications, as BSs acceptable delay for resource allocation in 5G networks is bounded to a few milliseconds in 3gpp recommendations. To address this issue, new resource allocation techniques \cite{Chen:2017:DSA, Sun:2017:EER, Sutton:2018:EUR, Ji:2018:URL, Kasgari: 2018: SOC} in 5G networks ensure the feasibility of URLLC by jointly considering the delay and reliability constraints while optimizing the resource allocation for BSs. For example, the authors in \cite{Chen:2017:DSA} proposed an optimal resource allocation strategy for uplink transmissions to maximize the delay-sensitive area spectral efficiency as a performance metric while guaranteeing the QoS constraint on reliability. In \cite{Sun:2017:EER} the authors proposed a method for maximizing energy efficiency for URLLC under strict QoS constraints on both end-to-end delay and overall packet loss. Sutton \emph{et~al.} \cite{Sutton:2018:EUR} investigated the potentials of using unlicensed spectrum for enabling ultra reliable and low latency communications, and suggested some promising use cases for that. Ji \emph{et~al.} \cite{Ji:2018:URL} discussed the physical layer challenges and requirements for URLLC and presented several enabling technologies for such communications in 5G NR downlink. Kasgari \emph{et~al.} \cite{Kasgari: 2018: SOC} presented a network slicing based resource allocation framework to provide reliable and low latency communications to users with such demands.\\ 

While the works in \cite{Chen:2017:DSA, Sun:2017:EER, Sutton:2018:EUR, Ji:2018:URL, Kasgari: 2018: SOC} discuss URLLC, they do not consider the joint user-to-BS association and resource optimization in this context. In this work, we first prove that the joint user association and resource allocation optimization in HetNets is a NP-hard problem which is intractable for large number of users. Then, we decompose the original optimization problem into a low complexity heuristic algorithm for user association and resource allocation. To further improve the time complexity of our heuristic method so as to make it attractive for URLLC, we use the Alternating Direction Method of Multipliers (ADMM) to solve the resource allocation optimization problem defined in our heuristic method. 

\section{System Model}
\label{sec3}

\subsection{Network Model}
We consider a HetNet model with one cellular base station at the center of the network who is responsible to ensure all cellular users will receive their required data service. We also have some overlaid small cell base stations which cover some of cellular users who located under their coverage range. Figure below show our network model, in which CBS uses high power transmissions (denoted with yellow links) to serve users who cannot be offloaded, while SBSs use low power transmissions (denoted with gray links) to serve cellular users under their coverage. 
\begin{figure}[htb!]
  \includegraphics[width=\linewidth]{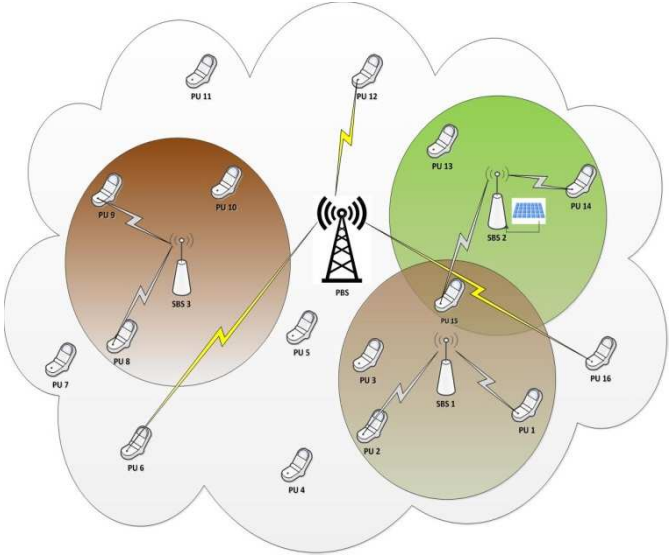}
  \centering
  \caption{Network Model.}
  \label{fig:boat1}
\end{figure}

\subsection{Auction Model}
In our model, we assume at the beginning of the offloading procedure, each SBS $k$ calculates its bandwidth requirements to serve each covered cellular user $i$, denoted as $\Phi_{k,i}$ which is a summation of two terms: $\Phi_{k,i}^p$ which is the required bandwidth to serve cellular user $i$, and $\Phi_{k,i}^s$ to serve its own users. The second term is the amount of reward that SBSs ask in exchange for serving cellular users under their coverage. So, the total amount of bandwidth asked by SBSs to serve any user $i$ is given by

\begin{equation}
\Phi_{k,i}=\Phi_{k,i}^p+ \Phi_{k,i}^s.
\end{equation}

Assuming the $k$-th SBS's transmit power-spectral density is $p^s$, and the channel fading between the $k$-th SBS and the $i$-th cellular user is denoted by $h_{k,i}^s$, the required bandwidth by SBS $k$ to serve cellular user $i$, $\Phi_{k,i}
^p$ can be derived by solving the equation

\begin{equation}
r_i^{min}=\Phi_{k,i}^p ~ log⁡(1+(p^s |h_{k,i}^s |^2)/N_0),
\end{equation} where, $r_i^{min}$ is the minimum data rate of cellular user $i$ that should be guaranteed by its serving base station. We also assume the overall cost of the cellular base station is given by

\begin{equation}
C=\sum \limits_{i\in U} \mu_i c_p p_i + c_w w_i,
\end{equation} in which, $\mu_i$ is the user association variable where, $\mu_i=0$ if user i is offloaded to SBSs, and  $\mu_i=1$ if user $i$ is not offloaded; $p_i$ is the transmission power spectral density of CBS while serving cellular user $i$; $w_i$ is the amount of bandwidth used by CBS to serve cellular user $i$; and $U$ is the set of cellular users.

\section{CBS Optimal Resource Allocation (ORA)}
\label{sec4}
After defining the cost function of cellular base station, we formulate and optimization problem to minimize CBS's cost, while guaranteeing the minimum data rate for cellular users. The ORA problem formulation is given by:

\begin{flalign}
&\min _{(\mu_i, w_i, p_i)} \sum \limits_{i\in U} (\mu_i c_p p_i + c_w w_i), &\\\nonumber\\
&~~~~~ subject~to: \nonumber &\\
&\sum \limits_{i\in U} (\mu_i w_i + \beta_{k,i} \phi_{k,i}) = W, &\label{eq5}\\
& w_i log⁡(1+ p_i |h_i^p |^2/N_0 ) \geq \mu_i r_i^{min},~\forall i\in U, &\label{eq6}\\
& 0 \leq p_i \leq p^{max},~\forall i\in U,&\label{eq7}\\
& 0 \leq w_i \leq w^{max},~\forall i\in U,&\label{eq8}\\
&\mu_i+\beta_{k,i}=1~\forall i\in U. \label{eq9}&
\end{flalign}

The constraint in Eq. \ref{eq5} ensures that the overall bandwidth used by cellular base station to serve users locally or to offload them to SBSs will not exceed the total available bandwidth, denoted as $W$. The constraint in Eq. \ref{eq6} guarantees that the data rate offered to cellular users will be higher than the minimum data rate required by them. Note that in the right hand side of this constraint $r_i^{min}$ is multiplied by the binary user association variable, $\mu_i$, which means that if user i is not been offloaded ($\mu_i=1$), then cellular base station has to guarantee $r_i^{min}$ data rate, otherwise if user i is offloaded ($\mu_i=0$), the right hand side becomes zero, meaning that cellular base station does not require to guarantee any data rate to this user, i.e. $p_i=0$, $w_i=0$. The constraints in Eq. \ref{eq7} and Eq. \ref{eq8} ensure that the allocated power and bandwidth to each user $i$ fall within the the feasibility regions for power and bandwidth variables, respectively.  And the last constraint in Eq. \ref{eq9} ensures that each user $i$ can be served by either cellular or a small cell base station, and not both of them. Note that $\beta_{k,i}$ is a binary association variable with $\beta_{k,i}=1$ is user $i$ is offloaded to SBS k, and $\beta_{k,i}=0$, otherwise, where $k$ is the index of the best serving SBS for each user $i$

\subsection{Complexity and Scalability issues of ORA}
In the initial ORA problem formulation presented in previous section, since it jointly optimizes all the user association, power and bandwidth allocation variables, it is a combinatorial problem with high time complexity \cite{Yousefvand:2017:DES}. In Theorem \ref{Theorem1} we prove that the ORA problem is NP-hard. Hence, its not scalable for large scale networks with large number of users as it cannot be solved in polynomial time; and also considering the dynamics of channel states in wireless networks which are varying by time, we need to find a low complexity solution to ORA problem to solve it within an acceptable time.

\newtheorem{thm}{Theorem}
\begin{thm}
\label{Theorem1}
The ORA problem is an NP-hard problem.\\

\textit {Proof:} We show that a simplified version of the ORA problem is reducible to the knapsack problem which is a well known NP-hard problem \cite{Garey:1979:CAI}; hence, ORA is also NP-hard. For the datails of the proof, see Appendix A. 
\end{thm} 

To address the scalability issue of the ORA we can replace it with a distributed user association and resource allocation method and use the techniques of game theory to solve it. For example, in \cite{Yousefvand:2018:IEU} we modeled the user association problem in HetNets in a form of a Stackelberg game between cellular and WiFi service providers as the leaders, and users as the followers of the game, and we found the Nash equilibria solutions to that problem under different conditions. The other solution to the scalability issue of ORA, which we propose in this work is to change the initial ORA formulation by breaking it down into two smaller size problems with low complexities. In this work we replace the initial ORA problem with a heuristic user association problem and a convex power/bandwidth allocation problem.

\section{Relaxed Heuristic Method (RHM)}
\label{sec5}
As mentioned in previous section, we replace the NP-hard ORA problem with a two phase low complexity solution, in  which we first solve the user association problem using a heuristic user association (HUA) algorithm, and then we optimize the CBS's power/bandwidth allocation using a Convex Resource Allocation (CRA) algorithm.

\subsection{Heuristic User Association (HUA)}
Since the user association variables in ORA optimization $(\mu_i, \forall i\in U)$ are binary variable and optimization problems with binary optimization variables are often NP-hard, in this section we propose a heuristic low complexity solution to user association problem which can be solved in polynomial time. From the Shannon formula for the capacity of wireless links we have 

\begin{equation}
r_i= w_i~log⁡(1+p_i |h_i |^2/N_0 ).
\label{eq10} 
\end{equation}

If we rewrite this formula in terms of transmit power we have
\begin{equation}
p_i= N_0( 2^{r_i/w_i} - 1 ) /|h_i |^2,
\label{eq11} 
\end{equation}
and taking a derivative of transmit power, $p_i$, with respect to data rate, $r_i$, we have

\begin{equation}
\frac {\partial p_i} {\partial r_i}= \frac{N_02^{r_i/w_i}ln⁡2}{w_i |h_i|^2} >0.
\label{eq12} 
\end{equation}

As we can see ,$\frac {\partial p_i} {\partial r_i}$, is always positive, meaning that CBS's required transmission power has a direct relation with the users offered data rate, and to minimize power consumption, base stations should minimize their offered data rate. So, it proves that the data rate constraints in ORA formulation given in Eq. \ref{eq6} will be always satisfied with equality, i.e. to minimize the power consumption, the cellular base station will always offer the minimum acceptable data rate to cellular users, $r_i=r_i^{min}$. Knowing that, we can replace the data rate variable $r_i$ with the minimum data rate $r_i^{min}$. In this case, the minimum required BW for cellular base station to serve any cellular user $i$, denoted as $w_i^{min}$ can be derived by solving 
\begin{equation}
r_i^{min}= w_i^{min}~log⁡(1+p^{max} |h_i^p |^2/N_0 ),
\label{eq13} 
\end{equation} where, $p^{max}$ is the maximum power spectral density of CBS, and $h_i^p$ is the channel gain between CBS and cellular user $i$.

After deriving the minim bandwidth required for serving all users, CBS  considers $w_i^{min}$, and $p^{max}$ as a rough estimations of the amount of bandwidth and power required to serve each user $i$, and uses this parameters to estimate the cost of serving for each user. Then, CBS compares the cost of serving with the cost of offloading for each user to make a decision about serving it directly or offloading it to SBSs. 
Assuming $c_w$ as the unit cost of bandwidth and $c_p$ as the unit cost of power for cellular base station, the serving cost of user $i$ can be estimated by $c_w w_i^{min}+ c_p p_i^{max}$. Also, the cost of offloading for user $i$ is given by $c_w \phi_{k^*,i}$, in which $k^*$ is the index of the best serving WiFi SP who asked for minimum amount of bandwidth to serve user $i$. After calculating the cost of offloading and the estimated cost of serving for each user, our heuristic algorithm will solve user association by simply comparing the offloading and serving costs for each user.

The proposed heuristic user association algorithm is given as follows:
\begin{algorithm}
\caption{Heuristic User Association (HUA)}
\label{HUA}
\begin{algorithmic}
\For{users i from 1 .. N}
\State $k^* \gets arg \min_k⁡ \phi_{k,i} $ \Comment{find the best serving WiFi BS}
\If{$(~c_w \phi_{k^*,i} \geq ( c_w w_i^{min}+ c_p p_i^{max}))$} 
\State Serve user $i$ locally: $\mu_i=1,~\beta_{k^*,i} =0$
\State \ElsIf{$(~c_w \phi_{k^*,i} < ( c_w w_i^{min}+ c_p p_i^{max}))$}
\State Offload user $i$ to WiFi BS $k^*$: $\mu_i=0,~\beta_{k^*,i}=1$
\Return {$\mu_i, \beta_{k^*, i}~\forall i$}
\EndIf
\EndFor

\end{algorithmic}
\end{algorithm}

By running this heuristic, cellular base station determines which users should be offloaded and which users should be served directly, to minimize its cost. After running this heuristic algorithm, the $\mu_i$ and $\beta_{k^*,i}$ are known parameters to CBS and we do not need to define them as variables in the optimization problem for resource allocation. It simplifies the initial ORA problem greatly by removing the binary optimization variables from it and replacing them with given parameters.

\subsection{Convex Resource Allocation (CRA)}
Knowing the values of user association variables, we can simplify the original ORA problem by reformulating it in a form of a convex optimization problem. To do that, we first use the convex cost function
\begin{equation}
C= c_p^T p+\gamma c_w^T w,
\label{eq14}
\end{equation}
to calculate the CBS's cost, which is a linear and convex function in both $p$, and $w$ variables. The first term of this cost function denotes the CBS's power consumption cost which is the multiplication of unit cost vector by the actual power allocation vector of CBS, p. Also, the second term in Eq. \ref{eq14} denotes the CBS's bandwidth cost regularized by a regularization parameter $\gamma$, $0< \gamma \leq 1$, which models the trade-off between power and bandwidth costs. The vectors $c_p$ and $c_w$ are the unit cost vectors for power and bandwidth consumption, respectively.
We rewrite the data rate constraint, given in Eq. \ref{eq6} of the original ORA problem, in a form of a power allocation constraint as
\begin{equation}
p_i \geq (e^{w_i \mu_i r_i^{min}} -1) N_0/|h_i^p |^2 , \forall i \in U.
\label{eq15}
\end{equation}

Since all the parameters in the right hand side of Eq. \ref{eq15} are given parameters except the $w_i$, which is the bandwidth allocation variable, we use a change of variable 
\begin{equation}
\sigma_i= (e^{w_i \mu_i r_i^{min} }-1) N_0/|h_i^p |^2 \label{eq16}  
\end{equation}
 where $\sigma_i$ is just a function of $w_i$, and if we know its value, then $w_i$ can be uniquely derived from it. Now, we can rewrite the data rate constraint as a power allocation constraint as $p_i \geq \sigma_i, \forall i \in U$. Now that we replace all the constraints in initial ORA problem with linear constraints, the Convex Resource Allocation (CRA) problem can be formulated as:

\begin{flalign}
& \min_{w,p}⁡ c_p^T p+ \gamma c_w^T w, \label{eq17}&\\
& ~~~~~~subject~to: \nonumber \\
& \mu ^T w \leq W, &\label{eq18}\\
& \mu ^T \sigma \leq p \leq p^{max},   \forall i\in U, &\label{eq19}\\
& 0\leq w_i \leq w^{max}, \forall i \in U, \label{eq20}&
\end{flalign} where, w is the bandwidth allocation vector, $p$ is the power allocation vector, and $\sigma$ is the vector of minimum transmission powers required by CBS to serve each user. In the left hand side vector of Eq. \ref{eq19}, $\mu_i \sigma_i$ is equal to $0$ if user $i$ is offloaded, and is equal to $\sigma_i$, otherwise. Note that from Eq. \ref{eq16}, for every given vector of w, $\sigma$ becomes a known parameter; hence, this CRA optimization problem turns into a linear problem (LP) in p. Similarly, for every given vector of $p$, the CRA problem will become an LP in $w$. We can use this fact to solve the CRA problem using iterative numerical methods like Alternating Direction Method of Multipliers (ADMM). 
Defining cost functions $f(p)= c_p^T p$, and $f(w)= \gamma c_w^T w$, the objective of the CRA problem is to minimize  $f(p) + f(w)=c_p^T p+ \gamma c_w^T w$ as the objective function. The optimal solution to such problem will be 
\begin{equation}
(p^*, w^*)=inf⁡\{f(p) + f(w) | r_i^{min}=w_i   log⁡(1+(p_i |h_i |^2)/N_0 ) \}.
\end{equation}

To use ADMM to achieve this solution, we need to write the dual version of the CRA problem. Since we want to minimize the distance between offered data rate $w_i log⁡(1+ p_i |h_i |^2/N_0 )$ and the required data rate $r_i^{min}$, in the Lagrangian function, $L_p$, we add a new variable $y$ which is equal to the difference between the offered and the required data rates, and is used to model the cost of deviating from target data rate. So, to use ADMM we use the Lagrangian function below for each user $i$ and for each given values of the variables $p_i$ and $w_i$, which are the amounts of power and bandwidth allocated to user $i$

\begin{equation}
L (p_i,w_i,y_i )=f(p_i) +f(w_i) + y_i(r_i^{min}-w_i   log⁡(1+ p_i |h_i |^2/N_0 ).
\end{equation}

Hence, the Lagrangian function considering all the users will be given as
\begin{equation}
L (p,w,y )=f(p) +f(w) + y(r^{min}-w   log⁡(1+ p |h |^2/N_0 ),
\end{equation} in which, $p$ and $w$ are power and bandwidth allocation vectors for all users in our network, and $r^{min}$ is a vector with the same dimensions which includes the minimum data rates for all users. We denote the values of power and bandwidth allocation variables at each step $k$ as $p^k$ and $w^k$, respectively. To minimize the Lagrangian function using an iterative updates, we start from two initial values for power and bandwidth variables. Then, at each step $k$ we use the values of these parameters $p^k$ and $w^k$ to find the best value for the variable $y$ to be used in the next step, $y^{k+1}$, by minimizing the lagrangian function with respect to $y$. Then, using the value of $w^k,y^{k+1}$ we find the value of $p^{k+1}$, and using the values of $p^{k+1},y^{k+1}$ in the Lagrangian function we find the best value of $w^{k+1}$. This update procedure is given in:

\begin{equation}
\begin{cases} 
y^{k+1}= & Arg  \min_{y}⁡{L (p^k,w^k, y )}, \\ 
p^{k+1}= & Arg  \min_{p}⁡{L (p,w^k, y^{k+1})}, \\ 
w^{k+1}= & Arg  \min_{w}⁡{L (p^{k+1},w, y^{k+1} )} .\\ 
\end{cases}
\end{equation}

We continue this iterative updates until the convergence, which is defined as the state in which the offered data rates are very close to the required data rates, and $p$ and $w$ have their lowest possible value. Note that since we removed the binary user association variables, now the Lagrangian function is continuous and differentiable with respect to all the variables $p$, $w$, and $y$. And we can simply take a derivative from the lagrangian function in each iteration to find its optimal value w.r.t any variable while the other two variables are given parameters.

\section{Simulation Results}
\label{sec6}
\begin{figure}
  \includegraphics[width=\linewidth]{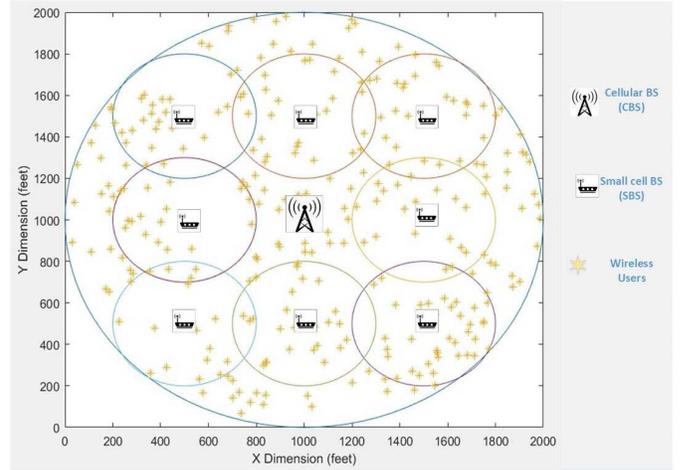}
  \centering
  \caption{Simulated Network Scenario.}
  \label{fig:simulatedscenario}
\end{figure}

To evaluate the efficiency of our proposed heuristic method, we consider a HetNet scenario in which there is one CBS located in the center of a cell with the radius of 1000 ft, and there are 8 overlaid SBSs with shorter coverage ranges of 300 ft within that cell who can serve the cellular users under their coverage. We also assume that there are 300 cellular users who are randomly distributed within the cell.
The simulated HetNet is shown in the Fig. \ref{fig:simulatedscenario}. For our first experiment, we assume that the minimum acceptable data rate for users, $b_{min}$, is equal to 128 kbps.

We solved the joint user association and resource allocation problem for the CBS using both ORA and RHM methods, by implementing these methods in Matlab. For better comparison, and to show the effects of users offloading on reducing the CBS's cost, we also implemented the Direct Serving Method (DSM) in which CBS serves all the cellular users directly without offloading any of them to SBSs. In DSM method CBS optimizes its bandwidth and power allocations to minimize its serving cost using the cost function defined in Eq. \ref{eq14}. We compare the performance parameters of these three methods in Table \ref{T1}. 

\begin{center}
\label{T1}
\begin{tabular}{|c|c|c|c| } 
 \hline
 Algorithm & DSM & ORA &  RHM	\\ [0.5 ex] 
  \hline\hline
 Runing Time (sec)& 53.2131 & 54.0355 &	3.6809 \\ 
 \hline
 Avg Cost Per User & 102.6990 & 68.6247 &	68.6456\\ 
 \hline
 Serving Rate & 100\% & 100\% &	100\%\\ 
 \hline
 Total Ofloaded Users & 0 & 226 &	226\\
 \hline
\end{tabular}
\end{center}

As we can see from this table, in both ORA and RHM methods 226 users (i.e. 75.33 \% of the total 300 users) have been offloaded to the overlaid SBSs. Offloading of these users reduces the CBS's average energy and bandwidth cost per user from 102.6990 unit cost (UC) in DSM method to 68.62 UC and 68.64 UC in ORA and RHM methods, respectively which leads to the reduction of CBS's total energy and bandwidth cost by nearly 33\% as compared to DSM. Also, as we can see in Table \ref{T1}, although the offloading rate and the average cost per user in both ORA and RHM methods are the same, but the RHM's running time is a fraction of the running time for ORA method. In fact, using the proposed heuristic method, we can reduce the resource allocation time for the CBS from 54.03 sec in ORA (which is an NP-hard method) to only 3.68 sec in RHM, which leads to the 93\% reduction in spectrum access delay for cellular users. For our simulations we used a laptop with an Intel core i5 2.3 GHz processor, and using a faster processor can further reduce the spectrum access delay to a few milliseconds.

\begin{figure}
  \includegraphics[width=\linewidth]{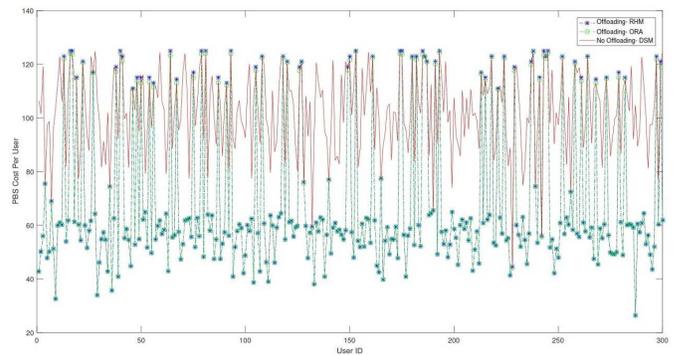}
  \centering
  \caption{Comparing cost per user in DSM, ORA and RHM.}
  \label{fig:cost comparison}
\end{figure}

Fig. \ref{fig:cost comparison} compares the CBS's energy and bandwidth consumption cost for each user in the ORA, RHM and DSM methods. As shown in this figure, for those users who have been offloaded by CBS using ORA method (which have service costs less than 80), the cost of serving is the same in both RHM and ORA methods which means that the RHM heuristic user association method has achieved the same results for user association as in ORA method. The reason is that the difference between the cost of serving and offloading for users who have been offloaded by ORA method is big enough that the heuristic user association method can also identify that offloading those users has lower cost than serving them directly. Also, it is shown in this figure that both RHM and ORA methods can reduce the CBS's average cost for the offloaded users significantly as compared the DSM method. The reason is that SBSs have better channel conditions and usually require less power and bandwidth to serve to serve cellular users under their coverage as compared to the CBS. So, by taking advantage of offloading users to SBSs, the CBS can reduce its energy and bandwidth consumption cost by 33\% as shown in the Table \ref{T1}.

\begin{figure}
  \includegraphics[width=\linewidth]{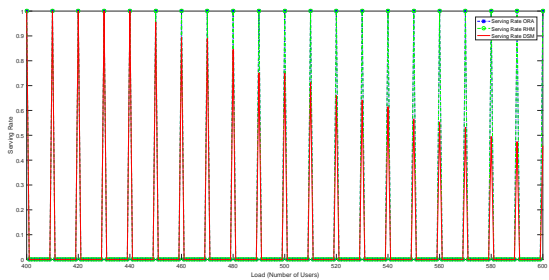}
  \centering
  \caption{Comparing Service Rates while increasing the load.}
  \label{fig:servicerates}
\end{figure}

To see the effects of load on the serving rate of cellular users in each of the ORA, RHM and DSM methods, we change the number of users in our HetNet model from 300 users to 500 users, by increasing it by 20 users in each step. We define the service rate as the percentage of users who are getting a data service with a rate higher than their minimum acceptable rate. The Fig. \ref{fig:servicerates} compares the service rates of DSM, ORA, and RHM methods under different load situations. As we can see, by increasing the load the CBS is unable to serve all the users in DSM method, and the service rate goes below 50\% if number of users exceeds 600, while in both ORA and RHM methods by exploiting the cooperation between CBS and WBSs, and offloading users to less congested small cells, the full service rate can be maintained despite the load.

Similar results are observed when we increase the minimum acceptable data rate for users, from 128 kbps to 512 kbps, to see its effects on the CBS's serving rate. Again, the service rates in DSM will diminish by increasing the minimum acceptable rate for users, while ORA and RHM methods can preserve the full service rate.

\section{Conclusion}
\label{sec7}
In this paper, we considered a HetNet model to optimize the user-to-BS association and resource allocation for cellular base stations while reducing the spectrum access delay for cellular users which is required for enabling URLLC in 5G. We first proposed an optimization method for joint user-to-BS association and resource allocation in HetNets. In order to reduce the high time complexity of the optimization method which makes it intractable for large number of users, we proposed a low complexity heuristic method which breaks down the initial optimization problem into a simple heuristic user association algorithm and a convex resource allocation optimization method. To further reduce the time complexity of the proposed solution, we proposed an ADMM-based iterative solution to the convex resource allocation problem. Simulation results validated the efficiency of the relaxed heuristic method, and showed that it can reduce the CBSs energy and bandwidth consumption by 33\%, while reducing the spectrum access delay for cellular users by 93\% which is attractive to URLLC. While in this paper, we used a HetNet model with only one CBS and multiple SBSs, the proposed approach is also applicable in HetNets with more than one CBS. In such a scenario, for each user, the CBS with the strongest radio link  (channel gain and/or RSSI) will serve as the primary CBS among all the candidate CBSs, and that CBS will make a  decision about serving each user directly or offloading it to overlaid SBSs to save energy and bandwidth following a similar procedure as outlined in the paper.

\appendices
\section{Proof of the Theorem 1}
\begin{proof}
We prove this theorem by showing that a simplified version of the ORA problem is reducible to the knapsack problem which is a well-known NP-hard problem \cite{Garey:1979:CAI}. As shown in Eq. \ref{eq11}, to serve any user $i$ associated to the CBS, i.e $\mu_i=1$, the CBS's transmit power is a function of its data rate and bandwidth. Assuming the amount of bandwidth allocated to user $i$, $w_i$, is fixed, the derivative of $p_i$ with respect to $r_i$ is given in Eq. \ref{eq12} where as we can see $\frac {\partial p_i} {\partial r_i}>0$. It means that by fixing the amount of bandwidth allocated to each user, $w_i$, the CBS transmit power increases by increasing the offered data rate to that user. Hence, the CBS will always serve the cellular users with the minimum data rate acceptable by them, $r_i^\text{min}$, to minimize its power consumption. So, we can simply remove the data rate constraint in Eq. \ref{eq6} from the ORA problem since it will always be satisfied with equality. For any user $i$ associated with CBS, the minimum bandwidth required by CBS to serve that user, $w_i^\text{min}$, can be derived from the Eq. \ref{eq13} when CBS is transmitting with its maximum power, i.e., $ p_i=p^\text{max}$.
So, by setting $ p_i=p^\text{max}$ and $w_i=w_i^\text{min}$ in the ORA problem, the constraints in Eq. \ref{eq7} and Eq. \ref{eq8} can be removed, and the simplified ORA problem can be rewritten as:
\begin{flalign}
&\min _{\mu_i} \sum \limits_{i\in U} (\mu_i c_p p^\text{max} + c_w w_i^\text{min}), &\\
&~~~~~ subject~to: \nonumber &\\
&\sum \limits_{i\in U} (\mu_i w_i^\text{min} + \beta_{k,i} \phi_{k,i}) = W, &\label{eq5}\\
&\mu_i+\beta_{k,i}=1~\forall i\in U. \label{eq9}&
\end{flalign}

Since by offloading each user $i$ to SBSs, CBS can save the amount of $p^\text{max}* w_i^\text{min}$ in transmit power, CBS should maximize the number of offloaded users using its available bandwidth, to minimize its cost. Now, if we define ${\rm{\Delta W}} = {\rm{W}} - \mathop \sum \limits_{{\rm{i}} \in {\rm{U}}} {\rm{w}}_{\rm{i}}^{{\rm{min}}}$ as the maximum amount of extra bandwidth at CBS that can be granted to SBSs as an incentive to serve cellular users, and ${\rm{\Delta }}{\phi _{{\rm{k}},{\rm{i}}}} = \max \left\{ {{\phi _{{\rm{k}},{\rm{i}}}} - {\rm{w}}_{\rm{i}}^{{\rm{min}}},{\rm{}}0} \right\}$ as an amount of extra BW required by best serving SBS to serve user $i$, then the simplified ORA problem can be formulated as:
\begin{equation}\label{eq166}
\mathop {\max }\limits_{\left( {{{\rm{\beta }}_{{\rm{k}},{\rm{i}}}}} \right)} \mathop \sum \limits_{{\rm{i}} \in {\rm{U}}} {{\rm{\beta }}_{{\rm{k}},{\rm{i}}}}{{\rm{p}}^{{\rm{max}}}}{\rm{w}}_{\rm{i}}^{{\rm{min}}}
\end{equation}

Subject to:
\begin{equation}\label{eq177}
\mathop \sum \limits_{{\rm{i}} \in {\rm{U}}} {{\rm{\beta }}_{{\rm{k}},{\rm{i}}}}{\rm{\Delta }}{\phi _{{\rm{k}},{\rm{i}}}} \le {\rm{\Delta W}}.
\end{equation} where $\beta_{k,i}$ is a binary association variable with $\beta_{k,i}=1$ is user $i$ is offloaded to SBS k, and $\beta_{k,i}=0$, otherwise, where $k$ is the index of the best serving SBS for each user $i$. The above problem is exactly equivalent to the zero-one knapsack problem which is an NP-hard problem. Hence, ORA problem is also NP-hard.
\end{proof}


\begin{IEEEbiography}[{\includegraphics[width=1in,height=1.25in,clip,keepaspectratio]{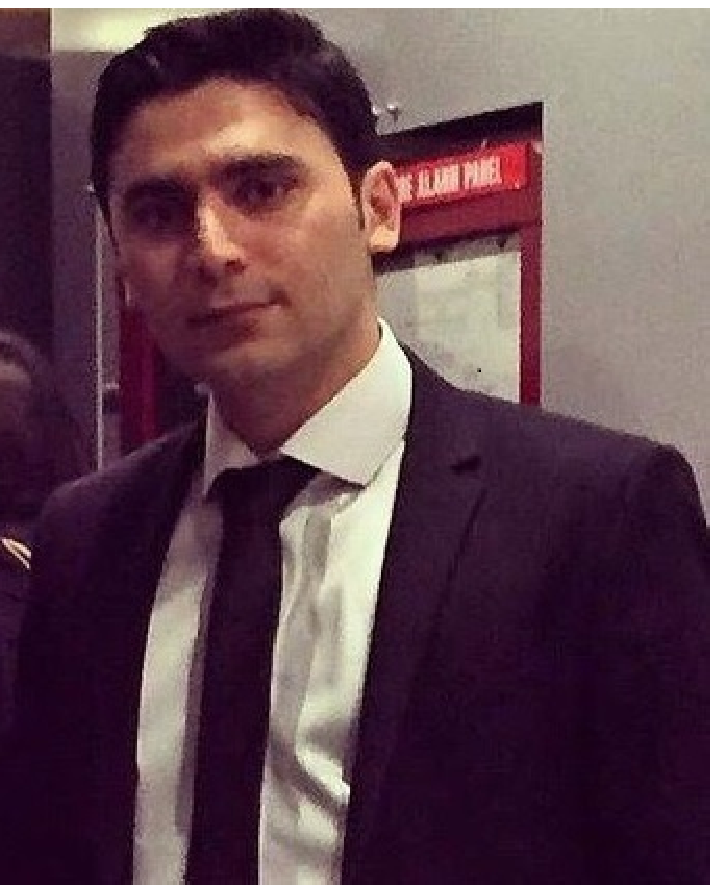}}]{Mohammad Yousefvand} is a phd candidate in Electrical Engineering in ECE department of Rutgers University, New Brunswick, NJ, US. He is currently with the Wireless Information Network Laboratory (WINLAB) at Rutgers University where he works on several NSF funded projects in different areas of wireless communications and networking. He is the recipient of Gary Thomas doctoral fellowship (annually $54,000$) from New Jersey Institute of Technology for 2014-2016 academic years, and professional development fund from Rutgers University for 2017-2018 academic years. His research interests include interference mitigation and resource allocation in wireless heterogeneous networks, cognitive radio, prospect theory, game theory, and performance evaluation.
\end{IEEEbiography}

\vskip 0pt plus -1fil

\begin{IEEEbiography}[{\includegraphics[width=1in,height=1.25 in,clip,keepaspectratio]{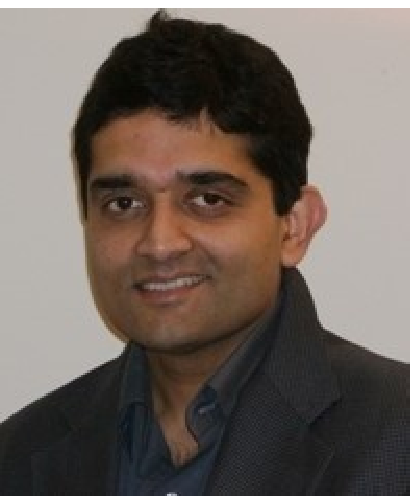}}]{Narayan B. Mandayam} ($S’89-M’94-SM’99-F’09$) received the B.Tech (Hons.) degree in 1989 from the Indian Institute of Technology, Kharagpur, and the M.S. and Ph.D. degrees in 1991 and 1994 from Rice University, all in electrical engineering.  Since 1994 he has been at Rutgers University where he is currently a Distinguished Professor and Chair of the Electrical and Computer Engineering department. He also serves as Associate Director at WINLAB. He was a visiting faculty fellow in the Department of Electrical Engineering, Princeton University, in 2002 and a visiting faculty at the Indian Institute of Science, Bangalore, India in 2003. Using constructs from game theory, communications and networking, his work has focused on system modeling, information processing and resource management for enabling cognitive wireless technologies to support various applications. He has been working recently on the use of prospect theory in understanding the psychophysics of pricing for wireless data networks as well as the smart grid.  His recent interests also include privacy in IoT, resilient smart cities as well as modeling and analysis of trustworthy knowledge creation on the internet.
Dr. Mandayam is a co-recipient of the 2015 IEEE Communications Society Advances in Communications Award for his seminal work on power control and pricing, the 2014 IEEE Donald G. Fink Award for his IEEE Proceedings paper titled “Frontiers of Wireless and Mobile Communications” and the 2009 Fred W. Ellersick Prize from the IEEE Communications Society for his work on dynamic spectrum access models and spectrum policy. He is also a recipient of the Peter D. Cherasia Faculty Scholar Award from Rutgers University (2010), the National Science Foundation CAREER Award (1998) and the Institute Silver Medal from the Indian Institute of Technology (1989). He is a coauthor of the books: Principles of Cognitive Radio (Cambridge University Press, 2012) and Wireless Networks: Multiuser Detection in Cross-Layer Design (Springer, 2004). He has served as an Editor for the journals IEEE Communication Letters and IEEE Transactions on Wireless Communications. He has also served as a guest editor of the IEEE JSAC Special Issues on Adaptive, Spectrum Agile and Cognitive Radio Networks (2007) and Game Theory in Communication Systems (2008). He is a Fellow and Distinguished Lecturer of the IEEE.
\end{IEEEbiography}


\end{document}